\documentclass{article}

\usepackage{apj}

\received{}
\accepted{}

\begin{document}

\title{Space VLBI Observations of 3C~371}

\author{Jos\'e-Luis G\'omez} 
\affil{Instituto de Astrof\'{\i}sica de Andaluc\'{\i}a, CSIC, Apartado 3004,
18080 Granada, Spain}
\authoremail{jlgomez@iaa.es}
\and
\author{Alan P. Marscher}
\affil{Department of Astronomy, Boston University, 725 Commonwealth
Avenue, Boston, MA 02215, USA}
\authoremail{marscher@bu.edu}

\begin{abstract}

  We present the first space VLBI observations of 3C~371, carried out at a
frequency of 4.8 GHz. The combination of the high resolution provided by the
orbiting antenna Highly Advanced Laboratory for Communications and Astronomy
(HALCA) and the high sensitivity of the VLBA allows imaging of the jet of
3C~371 with an angular resolution of approximately 0.26 mas, which for this
relatively nearby source corresponds to $\sim$ 0.4 h$^{-1}$ pc. Comparison
between two epochs separated by 66 days reveals no apparent motions in the
inner 7 mas jet structure above an upper limit of $\sim 1.4 h^{-1}$ c. This
value, the absence of detectable counterjet emission from the presumably
symmetric jet, plus the presence of extended double-lobe structure, are
consistent with the knots in the jet being stationary features such as
standing shocks. The jet intensity declines with the angular distance from the
core as $\phi^{-1.68}$. This is more gradual than that derived for 3C~120,
$\phi^{-1.86}$, for which there is evidence for strong intereactions between
the jet and ambient medium. This suggests that in 3C~371 there is a greater
level of {\it in situ} acceleration of electrons and amplification of magnetic
field. We interpret sharp bends in the jet at sites of off-center knots as
further evidence for the interaction between the jet and external medium,
which may also be responsible for the generation of standing recollimation
shocks. These recollimation shocks may be responsible for the presumably
stationary components. The radio properties of 3C~371 are intermediate between
those of other radio galaxies with bright cores and those of BL Lacertae
objects.


\end{abstract}

\section{Introduction}

  The radio source 3C~371 has been identified with an N galaxy by Wyndham
(\cite{W66}) at a redshift of 0.05 (Sandage \cite{S67}). The spectrum of this
object shows emission lines similar to those found normally in giant E
galaxies, but with a highly variable nonthermal continuum, which has led to
its classification as a BL Lacertae object (Miller \cite{Mi75}). 3C~371 has
been observed to be variable at a variety of wavelengths: radio (Aller, Aller,
\& Hughes \cite{AA92}), ultraviolet (Edelson \cite{E92}), optical (Carini,
Nobel, \& Miller \cite{C98}), and X-rays (Worrall et al. \cite{W84}).

  Very Long Baseline Interferometry (VLBI) observations of 3C~371 have shown a
complex milliarcsecond scale jet extending to the west (Pearson \& Readhead
\cite{PR81}, \cite{PR88}; Lind \cite{L87}; Polatidis et al. \cite{P95};
Kellerman et al. \cite{K98}). Polarimetric 5 GHz VLBI observations by Gabuzda
et al. (\cite{G89}) revealed very low polarization for the core ($\leq$0.3\%),
uncharacteristic of BL Lacertae objects. This, together with its low
bolometric luminosity (Impey et al. \cite{I84}), led them to suggest
that 3C~371 may be a transitional object between BL Lacertae and emission line
objects. Very Large Array (VLA) arcsecond resolution images (O'Dea, Barvainis,
\& Challis \cite{O88}; Wrobel \& Lind \cite{WL90}; Stanghellini et
al. \cite{S97}) show a partially bent jet with a magnetic field aligned with
the jet direction. High dynamic range 5 GHz VLA images by Wrobel \& Lind
(\cite{WL90}) have revealed twin lobes, a hot spot, and large radio halo,
unusual characteristics for a BL Lacertae object. An optical jet associated
with 3C~371 has also been reported by Nilsson et al. (\cite{N97}), who
detected a bright knot at 3'' from the nucleus, coincident with that observed
at radio wavelengths by Akujor et al. (\cite{A94}). Nilsson et al. interpreted
this as evidence for interaction between the jet and the external medium.

  We present here the first VLBI Space Observatory Program (VSOP) observations
of 3C~371. VSOP observations are performed using a ground array of radio
telescopes and the Japanese orbiting 8 meter antenna HALCA (Highly Advanced
Laboratory for Communications and Astronomy). The satellite has an orbital
period of approximately 6 hours, with apogee of 21000 km and perigee of 560
km. Data from HALCA are transmitted to a ground network of 5 tracking
stations, where they are recorded and subsequently correlated with the data
from the participating ground antennas.

\section{Observations and data reduction}

\begin{figure*}
\plottwo{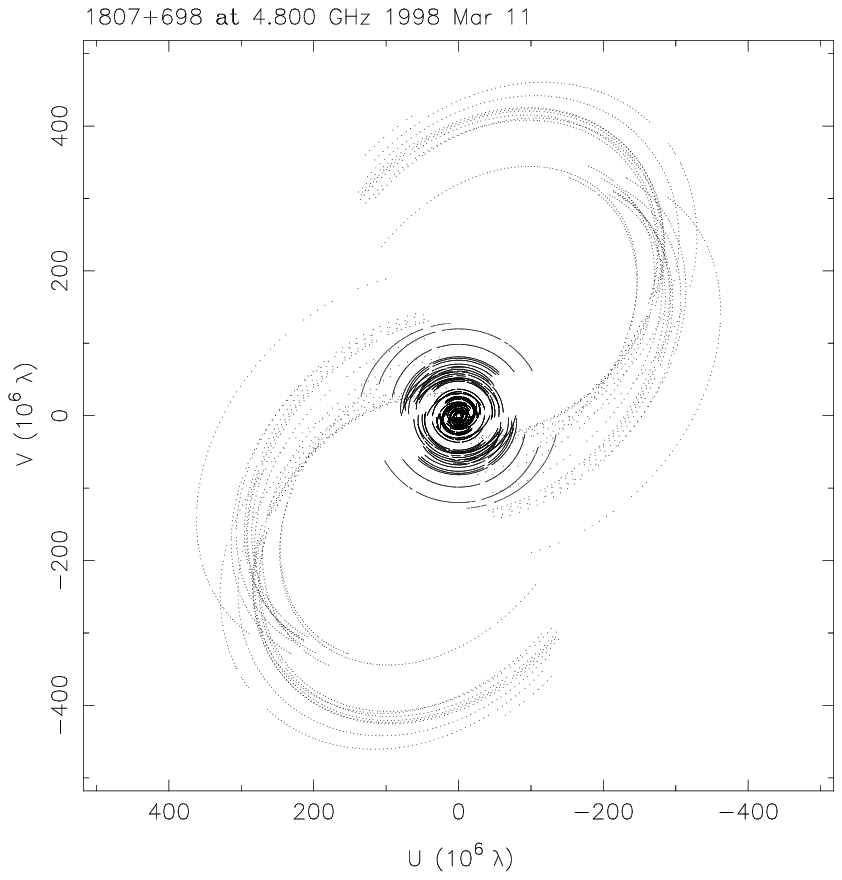}{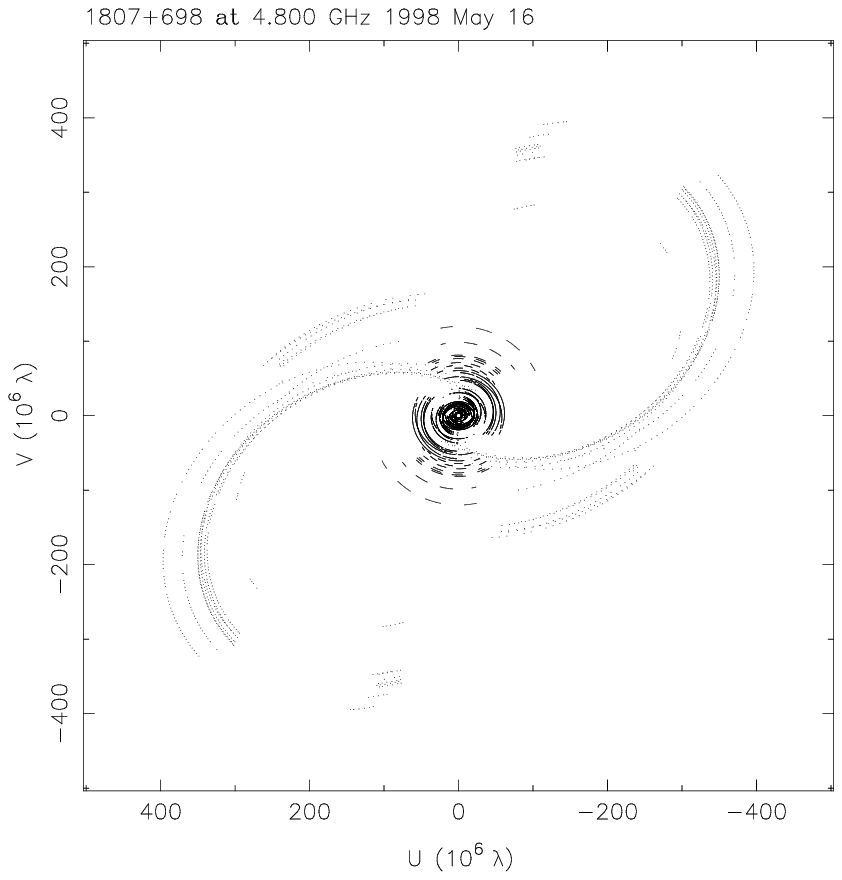}
%\vspace*{5cm}
\caption{The $uv$--coverage corresponding to VLBA+HALCA observations of 3C~371
on 1998 March 11 ({\it left}) and 1998 May 16 ({\it right}).}
\label{fig1}
\end{figure*}

  VSOP observations of 3C~371 were performed on 1998 March 11 (1998.19) and
1998 May 16 (1998.37) at a frequency of 4.8 GHz. NRAO's\footnote{The National
Radio Astronomy Observatory is a facility of the National Science Foundation
operated under cooperative agreement by Associated Universities, Inc.}  Very
Long Baseline Array (VLBA) provided the ground array for both observations
(with the exception of the St. Croix antenna, which did not participate in the
May observation). The data were recorded in VLBA format with two
intermediate-frequency bands (IF), centered on 4.800 and 4.816 GHz, each with
a bandwith of 16 MHz, and correlated in Socorro. The total durations of the
observations were 9.5 (March) and 8 (May) hr, each covering somewhat longer
than one HALCA orbit. Figure \ref{fig1} shows the {\it uv}--coverage obtained
for each epoch, where the baselines to HALCA are clearly visible, providing an
improvement in the resolution by more than 3 times that obtained using only
the ground telescopes. Since the observational resolution is proportional to
the frequency of observation, similar values can also be obtained by ground
arrays at shorter wavelengths.

  The reduction of the data was performed within the NRAO Astronomical Image
Processing System (AIPS) software. Data with very discrepant amplitudes and/or
phases on all baselines to a given antenna were deleted.  Fringe fitting was
performed to solve for the residual delay and fringe rate, resulting in good
solutions for most of the HALCA data. After averaging all frequency channels
the data were exported into the software DIFMAP (Shepherd 1997). An initial
time average of 10 seconds was introduced prior to phase calibration with a
point source model. A further time average of 2 minutes was introduced before
imaging.

  Imaging was performed through successive cleaning and self-calibration of
the data, both in phase and amplitude. In order to study the inner
milliarcsecond structure of 3C~371 with the highest resolution provided by
VSOP, we imaged the data using uniform weighting. The more extended structure
was revealed by making use of the ground data only, and with natural
weighting.

\section{Results}

\begin{figure*}
\plotone{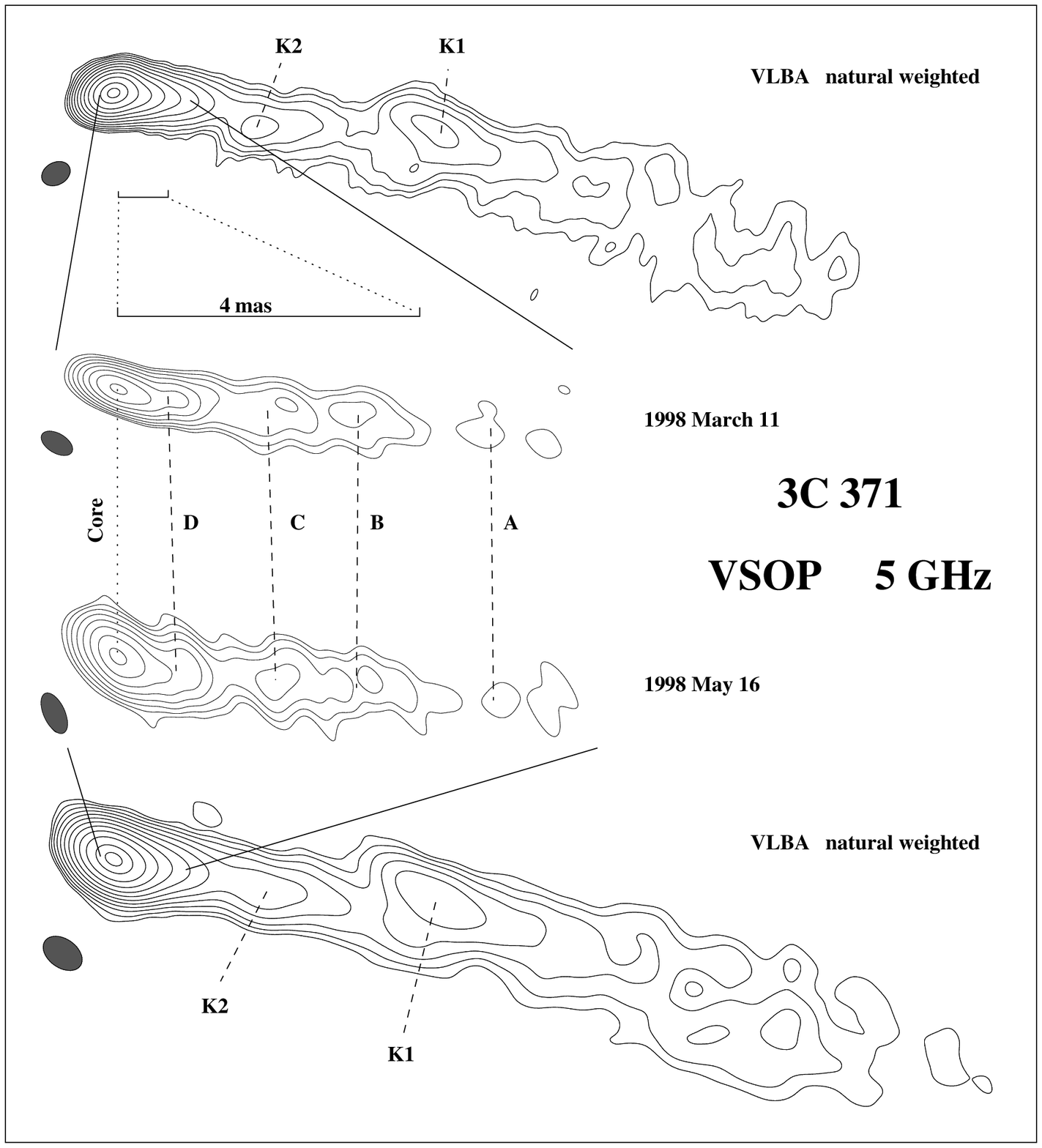}
%\vspace*{8cm}
\caption{VSOP total intensity images of 3C~371 at 5 GHz obtained on 1998 March
11 ({\it top middle image}) and 1998 May 16 ({\it bottom middle
image}). Natural weighted images obtained using only the ground array (VLBA)
data are shown, revealing the more extended jet structure. From top to bottom:
Contour levels are in factors of 2, starting at 0.06, 0.7, 0.8, and 0.06\%
(adding 90\% except for the March 1998 VSOP image) of the peak intensity
of 625, 331, 283, and 674 mJy beam$^{-1}$. Convolving beams (shown as filled
ellipses) are 2.4$\times$1.79, 0.457$\times$0.257, 0.576$\times$0.279, and
3.45$\times$2.3 mas, with position angles of 59$^{\circ}$, -62$^{\circ}$,
25$^{\circ}$, and 55$^{\circ}$.}
\label{fig2}
\end{figure*}

  Figure \ref{fig2} shows the uniformly weighted VSOP and naturally weighted
VLBA images obtained for 3C~371 at 1998 March 11 and 1998 May 16.  Components
in the total intensity images were analyzed by model fitting the {\it uv} data
with elliptical Gaussian components within DIFMAP. Because of the smooth jet
structure and non uniformity of the $uv$--coverage (see Fig. \ref{fig1}),
model fitting was rather complicated. Owing to the difficulty of using model
fitting to reproduce both very compact and more extended structure, we adopted
the following approach. Starting with the hybrid map delta-function model, we
deleted components within 7 mas of the core and then reconstructed the inner
jet by model fitting with elliptical Gaussian components.  Because of the
complex structure of knots {\it K1} and {\it K2} in the outer region, model
fitting of these components was accomplished by removing the delta-function
components within the highest contour levels surrounding their respective
positions, then model fitting the residuals with single elliptical Gaussians
for each component. We have estimated the positional errors in the model
fitting by introducing small changes in the position and studying the
variations in the $\chi^2$ of the fit, as well as by reproducing the model
fitting from the beginning and comparing with previous results. This yielded
uncertainties of $\sim$0.05 mas for the inner components {\it A} through {\it
D}, and $\sim$ 0.1 mas for knots {\it K1} and {\it K2}. Table 1 summarizes the
physical parameters obtained for 3C~371 at both epochs. Tabulated data
correspond to flux density ($S$), separation ($r$) and structural position
angle ($\theta$) relative to the core, FWHM major axis of the ellipical
Gaussian component ($a$), ratio of minor to major axis ($b/a$), and position
angle of the major axis ($\Phi$).

\subsection{Proper Motions}

  The possible existence of superluminal motions in the jet of 3C~371 is
controversial owing to the different values estimated by several
authors. Worrall et al. (\cite{W84}) fit the total millimeter to X-ray
spectrum of 3C~371 to the synchrotron-self-Compton jet model of K\"onigl
(\cite{K81}), obtaining a lower limit of 10 for the Doppler beaming factor,
$\delta_{\rm min}=1/[\Gamma(1-\beta\cos\theta)]$, where $\beta$ is the bulk
flow speed of the emitting plasma in units of $c$, $\Gamma =
(1-\beta^2)^{-1/2}$, and $\theta$ is the angle between the jet axis and the
line of sight. Under the assumption that $\sin\theta=1/\Gamma$, they predicted
superluminal motion with values larger than 10 $h^{-1}$c ($H_{\circ}$= 100 $h$
km s$^{-1}$ Mpc $^{-1}$). Similar values were suggested by Lind (\cite{L87})
through the analysis of two 5 GHz VLBI images obtained in 1982 and 1985. A
study of the changes in brightness distribution led these authors to conclude
that the apparent speeds were between 6 and 7 $h^{-1}$c.

  However, VLA observations by Wrobel \& Lind (\cite{WL90}) show that the
morphology of the extended radio emission of 3C~371 is that of a double-lobed
radio source. Since the lobes do not overlap, the viewing angle cannot be
extremely small. On the other hand, they detected no counterjet, which
suggests that the source does not lie near the plane of the sky. It is
therefore difficult to obtain an agreement between the large viewing angle
estimated by Wrobel \& Lind (\cite{WL90}), and the large apparent motions of
Worrall et al. (\cite{W84}) and Lind (\cite{L87}), since apparent motions of
the order of 10 would require a viewing angle smaller than approximately
$6^{\circ}$. Thus, a direct determination of the apparent proper motion is of
special importance for interpreting this source.

  The VLBA natural-weighted images of Fig. \ref{fig2} show a jet extending to
the west-southwest over about 60 mas. Superimposed on the homogeneous jet
brightness distribution, knots {\it K1} and {\it K2} appear at $\sim$ 12 and
26 mas from the core. The computed apparent motions of 1.2 and 0.3 mas/yr (2.8
and 0.7 h$^{-1}$c) for {\it K1} and {\it K2}, respectively, lie within the
estimated errors. We therefore obtain an upper limit to the apparent velocity
of the large scale jet structure of $\sim$ 3 h$^{-1}$c. Comparison with
previous 5 GHz VLBI maps by Lind (\cite{L87}) obtained in 1982 and 1985
reveals a tentative identification of {\it K1} and {\it K2} with their
components {\it C} and {\it E}, respectively, in which case both components
would have remained stationary over a 15-yr period, in discrepancy with the
value of 7 h$^{-1}$c estimated by Lind (\cite{L87}).

  The VSOP images of Fig. \ref{fig2} reveal up to four distinct components,
mapped with an angular resolution of $\sim$ 0.26 mas, which for this nearby
source corresponds to a linear resolution of $\sim$ 0.4 h$^{-1}$ pc. The
apparent velocities for components {\it A} through {\it D} range between 0.4
and 1.2 $h^{-1}$ c. However, the relatively large errors in the model fitting
of this complex source (see also Gabuzda et al. \cite{G89}) lead to an
estimated error in the apparent motion of $\sim 1.4 h^{-1}$ c, and therefore
we can only conclude that any motions in the inner structure of 3C~371 must be
smaller than this uncertainty.

\begin{figure*}
\plotone{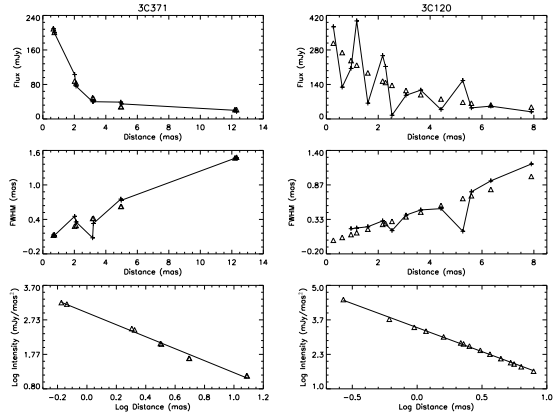}
%\vspace*{2cm}
\caption{Flux density ({\it top}) and FWHM angular size ({\it middle})
vs. angular distance of the components in 3C~371 ({\it left}) and 3C~120 ({\it
right}, from G\'omez, Marscher, \& Alberdi \cite{JL99}). Data for component
{\it K1} is neglected since the component does not fill the entire jet
width. Sizes of components in 3C~120 for which no estimation of the FWHM have
been obtained are ignored. The data are represented by crosses.  Triangles
correspond to the fitted exponential decay for the flux density, and linear
fit for the FWHM. Jet intensity decays with angular distance as $\phi^{-1.68}$
for 3C~371 and $\phi^{-1.86}$ for 3C~120 ({\it bottom} plots).}
\label{fig3}
\end{figure*} 

\section{Discussion and Conclusions}

  Our rather low upper limit to the apparent velocity of the inner jet
requires that either (1) the Lorentz factor is low, (2) the viewing angle is
large, (3) the Lorentz factor is high but the viewing angle is small, $\theta
\ll \Gamma^{-1}$, or (4) the knots in the jet are stationary features. We
assume that the absence of counterjet emission in the images of
Fig. \ref{fig2} is due to beaming of an otherwise symmetric twin jet, and
estimate a minimum observed jet:counterjet emission ratio of $\sim 1700:1$
(peak flux:noise level). In the case of {\it optically thin} emission in a
steady-state jet with randomly oriented magnetic field, this ratio is given by
$[(1+\beta\cos\theta)/(1-\beta\cos\theta)]^{2-\alpha}$, where $\alpha$ is the
spectral index ($S_{\nu}\propto\nu^{\alpha}$); $\alpha=-0.03$ in 3C~371
(Taylor et al. \cite{T96}). We note that for jet regions with relatively large
opacity, such as the brightest region of the radio core, the above equation is
not valid, and the jet:counterjet flux ratio is significantly smaller than
that obtained using the previous equation. Therefore, the use of this equation
provides a lower limit to the flow Lorentz factor, and a corresponding upper
limit to the viewing angle. Solving the jet:counterjet flux ratio equation
gives $\beta\cos\theta > 0.95$, or $\beta\ge 0.95$ ($\Gamma\ge 3.2$) and
$\theta\le 18^{\circ}$. Hence, if the jet is symmetric, we can eliminate
possibilities (1) and (2). Option (3) seems unlikely, since the twin-lobe
extended radio structure (Wrobel \& Lind \cite{WL90}) is not easily explained
if the viewing angle is much less than $\theta \sim 18^{\circ}$. Hence, we
conclude that option (4) is the most plausible, especially if we consider the
previously mentioned opacity effects in the determination of the
jet:counterjet flux ratio. Further strong evidence supporting the idea that
the components may be stationary features comes from the comparison with
images of similar resolution obtained by Kellerman et al. (\cite{K98}) using
the VLBA at 15 GHz (see http://www.cv.nrao.edu/2cmsurvey). These images show a
knotty jet with only minor structural changes during the four years covered by
their observations, and resembles the structure of Fig. \ref{fig2}.

  Similar to the radio galaxy 3C~120 (Walker \cite{W97}; G\'omez et
al. \cite{JL98}), the jet of 3C~371 displays a gentle curvature. Sharper bends
are observed farther downstream of the emission shown in Fig. \ref{fig2}
(Wrobel \& Lind \cite{WL90}; Akujor et al. \cite{A94}); these have been
interpreted as the interaction of the jet with the external medium (Nilsson et
al. \cite{N97}). The fact that knots {\it K1} and {\it K2} are offset with
respect to the jet ridge line supports this interpretation. They may arise
from Kelvin-Helmholtz instabilities (as seems to be the case in 0836+710;
Lobanov et al. \cite{L98}), or perhaps some other consequence of strong
interactions with the external medium (cf. 3C~120; G\'omez, Marscher, \&
Alberdi \cite{JL99}). If the jet of 3C~371 is confined by the pressure of the
external medium, the slow motion (consistent with no motion) observed in
components {\it D} through {\it A} suggests that these components may be
associated with recollimation shocks. Numerical simulations by G\'omez et
al. (\cite{JL95}, \cite{JL97}) show that the enhancement of specific internal
energy and rest-mass density at internal oblique shocks results in the
appearance of a regular pattern of knots of high emission. The separation and
strength of the knots is a function of the jet opening angle and Mach number.

\begin{tiny}
\begin{deluxetable}{lcccccccccccc}
\tablecolumns{13}
\tablewidth{0pc}
\tablecaption{Physical Parameters of 3C~371 \label{tab1}}
\tablehead{
Comp.&
\multicolumn{2}{c}{$S$}&
\multicolumn{2}{c}{$r$}&
\multicolumn{2}{c}{$\theta$}&
\multicolumn{2}{c}{$a$}&
\multicolumn{2}{c}{$b/a$}&
\multicolumn{2}{c}{$\Phi$}\nl
 & 
\multicolumn{2}{c}{(mJy)}&
\multicolumn{2}{c}{(mas)}&
\multicolumn{2}{c}{($^{\circ}$)}&
\multicolumn{2}{c}{(mas)}&
\multicolumn{2}{c}{}&
\multicolumn{2}{c}{($^{\circ}$)}}
\startdata
&1998.19&1998.37&1998.19&1998.37&1998.19&1998.37&1998.19&1998.37&1998.19&1998.37&1998.19&1998.37\nl
\hline
Core\dotfill&444&469&... &... &... &... &0.38&0.45&0.17&0.18& 75& 78\nl
D\dotfill   &209&200&0.67&0.73&-102&-104&0.61&0.69&0.27&0.22&-87&-85\nl
C\dotfill   &103& 76&2.02&2.11& -99& -98&1.11&0.72&0.52&0.63& 83&-70\nl
B\dotfill   & 40& 40&3.21&3.16& -97& -97&0.79&1.01&0.53&0.10& 41&-84\nl
A\dotfill   & 39& 35&4.95&4.98& -97& -97&2.34&2.33&0.41&0.40& 82& 82\nl
K2\dotfill  & 20& 17&12.2&12.3&-103&-102&3.27&2.75&0.57&0.68& 88& 62\nl
K1\dotfill  & 15& 16&26.5&26.4& -97& -97&2.76&2.41&0.57&0.75& 57& 69\nl
\enddata
\end{deluxetable}
\end{tiny}

  The images allow us to measure the gradient in intensity of the jet. For
that we have represented in Fig. \ref{fig3} the flux density and the FWHM
angular size of the components in 3C~371, for both epochs, as a function of
the distance along the jet. The flux density evolution can be fitted to an
exponential function of exponent -0.77, while for the component sizes a linear
fit gives a jet opening half-angle of $3.3^{\circ}$. We have neglected the
data corresponding to component {\it K1}, since at this position the jet seems
considerable wider than its estimated FWHM. Using these fits, we have
interpolated the flux and jet width at the position of the components,
obtaining a jet intensity which decays with angular distance $\phi$ from the
core as $\phi^{-1.68}$, shown in Fig. \ref{fig3}. For comparison, we have
performed a similar analysis for the jet in 3C120 using the data in G\'omez,
Marscher, \& Alberdi (\cite{JL99}). In this case, the flux decays with an
exponent of -0.45, and the jet half opening angle is $3.7^{\circ}$. The
intensity is found to decay as $\phi^{-1.86}$. The flatter gradient in the
case of 3C~371 suggests that in this object there is a greater level of {\it
in situ} acceleration of electrons and amplification of magnetic field than
for 3C~120, for which G\'omez, Marscher, \& Alberdi (\cite{JL99}) found
evidence of strong interactions between the jet and ambient medium. This would
lead to the physical conditions (high energy electrons and acceleration)
necessary in order to explain the optical jet observed in this source (Nilsson
et al. \cite{N97}). It would also imply that the jet of 3C~371 undergoes
greater interaction with the external medium, which is a likely source of the
shock waves and turbulence that could cause this amplification. As discussed
above, such interactions with the external medium would also excite standing
shocks in the jet, leading to the appearance of stationary components.

  The properties of 3C~371 indicate that this source may be a hybrid of a BL
Lac object and a radio galaxy. That is, it may represent a BL Lac object
observed at a somewhat wider viewing angle than is normally the case.  The
radio galaxy aspects of such an object should then become at least partly
apparent, as indeed occurs in 3C~371. It would be of great interest to make
higher frequency VLBI observations of this source, especially with
polarization, in order to investigate possible motions closer to the central
engine and to test whether indeed 3C~371 can be considered a true transitional
object.

\begin{acknowledgements}
This research was supported in part by Spain's Direcci\'on General de
Investigaci\'on Cient\'{\i}fica y T\'ecnica (DGICYT) grant PB97-1164, by NATO
travel grant SA.5-2-03 (CRG/961228), and by U.S. National Science Foundation
grant AST-9802941.
\end{acknowledgements}

\end{document}